\documentclass[journal]{IEEEtran}
\ifCLASSINFOpdf
\else
\fi
\usepackage{graphicx}

\begin{document}
%
\title{Machine Ruling}
%
%
%

\author{Zhe~Chen
\thanks{Zhe Chen is currently with 
Northeastern University, China,
 e-mail: zchen42@students.tntech.edu, chenzhe@ise.neu.edu.cn, zhechen2008@gmail.com.}
}

%
%

\markboth{Chen: Machine Ruling,~2015}%
{Chen: Machine Ruling}
%



\maketitle

\begin{abstract}
Emerging technologies, such as big data, Internet of things, cloud computing, mobile Internet, and robotics, breed and expedite new applications and fields.
In the mean while, the long-term prosperity and happiness of human race demands 
advanced technologies.
In this paper, the aforementioned emerging technologies are applied to management and governance for the long-term prosperity and happiness of human race. The term ``machine ruling'' is coined, introduced, and justified. Moreover, the framework and architecture of machine ruling are proposed. Enabling technologies and challenges are discussed.

\end{abstract}

\begin{IEEEkeywords}
Machine Ruling, Big Data, Internet of Things, Cloud Computing, Management, Governance.
\end{IEEEkeywords}

%
\IEEEpeerreviewmaketitle

\section{Introduction}
%
%
%
%

 %
%


\IEEEPARstart{S}{peaking} of machine and ruling, people may remind of the Skynet, an artificial general intelligence seeking to destroy the human race, in the film \textit{Terminator Genisys}. Many fictions and films depict the war between human race and intelligent machines. Would machines eventually replace human beings on this planet? Not really. People who fear intelligent machines misunderstand what computers are doing when they say machines are thinking or getting smart~\cite{mit_tech_rev}. 
Intelligent machines or computers can do better in terms of computation and memorization than human beings. However, they are not able to think in the way that human beings do, hence they are not smart enough actually.

Human beings use tools to a much higher degree than any other animals. 
A machine is no more than a tool containing one or more parts that use energy to perform an intended action.
Human beings harness machines to serve themselves. 
In this paper, machines generally refer to those devices that perform sensing, communicating, storing, computing, and executing tasks. 

With the rapid advance of information technology, the world has changed a lot. Today, widely used smart phones enable people to access the Internet at anywhere and anytime with enhanced computation and storage capability. More and more objects are connected to the Internet. The way of life is changing. 
However, so far the advance of technology does not necessarily lead to a peace and better life. The terror happened in Paris in November, 2015,  shocked the world. 
At the same time, a major part of the wealth of the world is being transferred to a minor part of human beings. The rich get richer and the poor get poorer, which makes the world getting greedy and conflicted. Although human natures are unblamable,
for the long-term prosperity and happiness of human race, 
the whole world needs to work together to achieve this common goal.

But how? Some measures must be taken in order to achieve this common goal. Since machines are good at computation and memorization and they do not possess human weaknesses, such as greed, fear, selfishness, deceit, and myopia, machines are ideal to assist human beings in management and governance.

In this paper, the term ``machine ruling'' is coined. The framework and architecture for machine ruling are also proposed. 

The paper is organized as follows. Section II specifies the reasons for machine ruling. Section III introduces the concept of machine ruling. Section IV proposes a framework and an architecture for machine ruling. Section V discusses the enabling technologies and challenges for implementing machine ruling. And Section VI concludes this paper.

\section{Why Is Machine Ruling}

Why is machine ruling necessary? Will it prevail?

First, machines can make fair, optimized, and reliable decisions. Human beings are often affected by emotions and other subjective factors when making decisions. Moreover, human beings usually have limited information for making decisions, since it is very hard for human beings to exactly remember everything they learned and experienced and to know everything directly or indirectly relevant to the decisions. 

With human governance, the future prosperity of a territory much depends on the characteristics of major leaders, especially in developing countries. 
In democratic countries, although the power of governance is balanced in some sense, the efficiency of governance is debatable. In non-democratic countries, the risk of dictatorship is a major concern. 
As a Chinese saying goes, ``ren wu wan ren'', which means no one is one hundred percent perfect. To reduce the uncertainty and risk of choosing unsuitable leaders, machine ruling is a better choice. 

In many countries, some important information is deliberately concealed to the public, for the sake of preserving the governance or the interest of a small group of people. This is information unfairness. Moreover, for different groups of people, policies may differ. This is policy unfairness. Even the same policy may be explained and implemented in different ways for different groups of people. This is implementation unfairness. Machine ruling is proposed to deal with the information unfairness and the policy unfairness, and to partially deal with the implementation unfairness.
Machines are stable, reliable, and honest. 
Machines can make globally optimized decisions effectively, based on comprehensive information much more than human beings can deal with, for the long-term prosperity and happiness of human race, instead of for the interest of just a small group of people. Corruptions, absurdness, and moral degeneration, will not happen on machines. World wars or economic crises may not happen, either, if the world is managed by inter-connected machines.

Second, with machine ruling, the organizational structure of the world can be reduced to flat organization. A human manager can usually supervise several to tens of subordinates. Hence, hierarchical organization is necessary in current governance. 

Machines can take advantage of parallel computing and cloud computing for making decisions or assigning tasks to many people simultaneously. As long as computation and storage resources are sufficient, inter-connected machines are virtually able to serve all the people in the world. 
In such a flat organization, people are equally treated. Man-made discrimination and abuse of power will be eliminated. People just do what he or she is good at to contribute to the society and enjoy life, leaving alone all the management and coordination matters to machines. Since there is no intermediate management levels in flat organization, the efficiency of machine ruling can be very high.

Third, machines can be literately immortal. Human beings are mortal. People spend roughly one third of their lifetimes in growing up and studying to get prepared, and spend another one third of their lifetimes in working. People die with their knowledge and experiences, whereas newborns start learning everything from scratch. Although machines also get aged or damaged, their memories, where their knowledge and experiences are stored, can be exactly copied and backed up. So their knowledge and experience can be preserved and transferred to newly-built machines. In this sense, machines are immortal. 
Machines can accumulate and preserve knowledge and experiences, and knowledge and experiences are vital for management, which makes them suitable for managing the world with their long-standing extensive knowledge and experiences.

Fourthly, machines are designed to assist human beings. If there is anything that machines can do for humans, why not let them do? Management or governance is just one of these things. Since life is limited and precious, human beings just need to do what they are good at and what machines can not do. For the rest of their lifetimes, why not just enjoy life? Let machines worry about everything that machines can do. Imagine the following scenario. When you wake up, your home machines, or robots, bring you manchine-cooked healthy breakfast. When you enjoy your breakfast, machines talk with you about latest news that you are interested in and your daily schedule according to your status and your desire. When you step out of your home, an auto-driving transportation vehicle scheduled by machines is waiting for you. The route has already been optimized for you.
When you get to workplace, machines suggest you what to do and how to do. You just need to finish the work step by step according to the prompts given by machines. Machines automatically provide you with assistance whenever they figure out that you may need help. You do not need to take trouble to remember details, such as your accounts, passwords, files in your hard-drives, and whatever. Machines take care of all these matters for you. During your work time, machines clean your home and deliver supplies to your home. What a lovely day! 

Fifthly, the advance of technology is driving machine ruling possible. Recently, enabling technologies, such as Internet of things, mobile Internet, big data, cloud computing, and robotics, are emerging and developing fast. These technologies cover sensing, communicating, storing, computing, and executing, providing machine ruling with a comprehensive systematic support.

The age of machine ruling will eventually come.

\section{What Is Machine Ruling}

The term ``ruling'' here generally refers to management and governance. ``Machine ruling'' means to employ machines, including any sensing, communicating, storing, computing, and executing devices, to perform management functions, such as planning, organizing, leading, and controlling~\cite{management}, and to perform governance functions, such as determining objectives, determining ethics, creating culture, ensuring compliance, ensuring accountability, and implementing governance framework~\cite{governance}, 
for the long-term prosperity and happiness of human race.

The key to machine ruling is that machines have the abilities to judge, to infer, and to learn from massive data, as well as to recover, to defend, and to evolve themselves.

\section{Proposed Framework and Architecture of Machine Ruling}

\begin{figure*}[!t]
	\centering
	\includegraphics[width=5.5in]{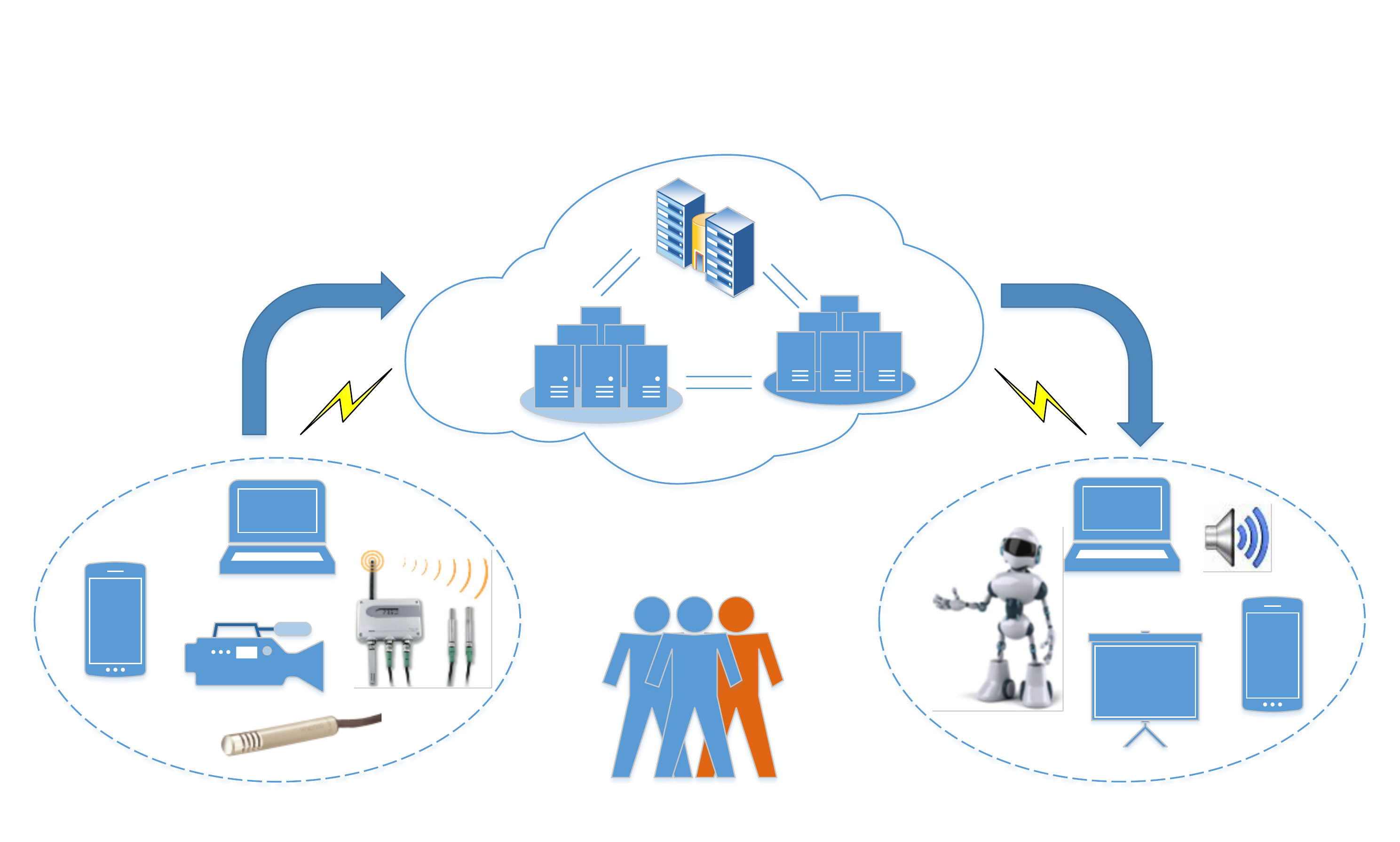}
	\caption{Proposed framework of machine ruling.}
	\label{fig_framework}
\end{figure*}

Fig.~\ref{fig_framework} depicts the proposed framework for machine ruling. Input devices, including personal computers or laptops, smart phones, cameras, and active and passive sensors, gather all kinds of information about the physical world and human beings, such as positions, videos, voices, and ambient parameters. The gathered information is digitized and transmitted to data centers in the cloud by either wireline or wireless. Data centers in the cloud store and backup the received data, and run software and algorithms to react to the received data. Processing results and instructions are then transmitted to output devices, including personal computers or laptops, smart phones, screens, speakers, and robots, again by either wireline or wireless. The output devices then execute the given instructions or feedback the processing results to human beings. What human beings react to the results will be caught by input devices and transmitted to data centers. All these steps constitute an interactive cycle.

\begin{figure*}[!t]
	\centering
	\includegraphics[width=7in]{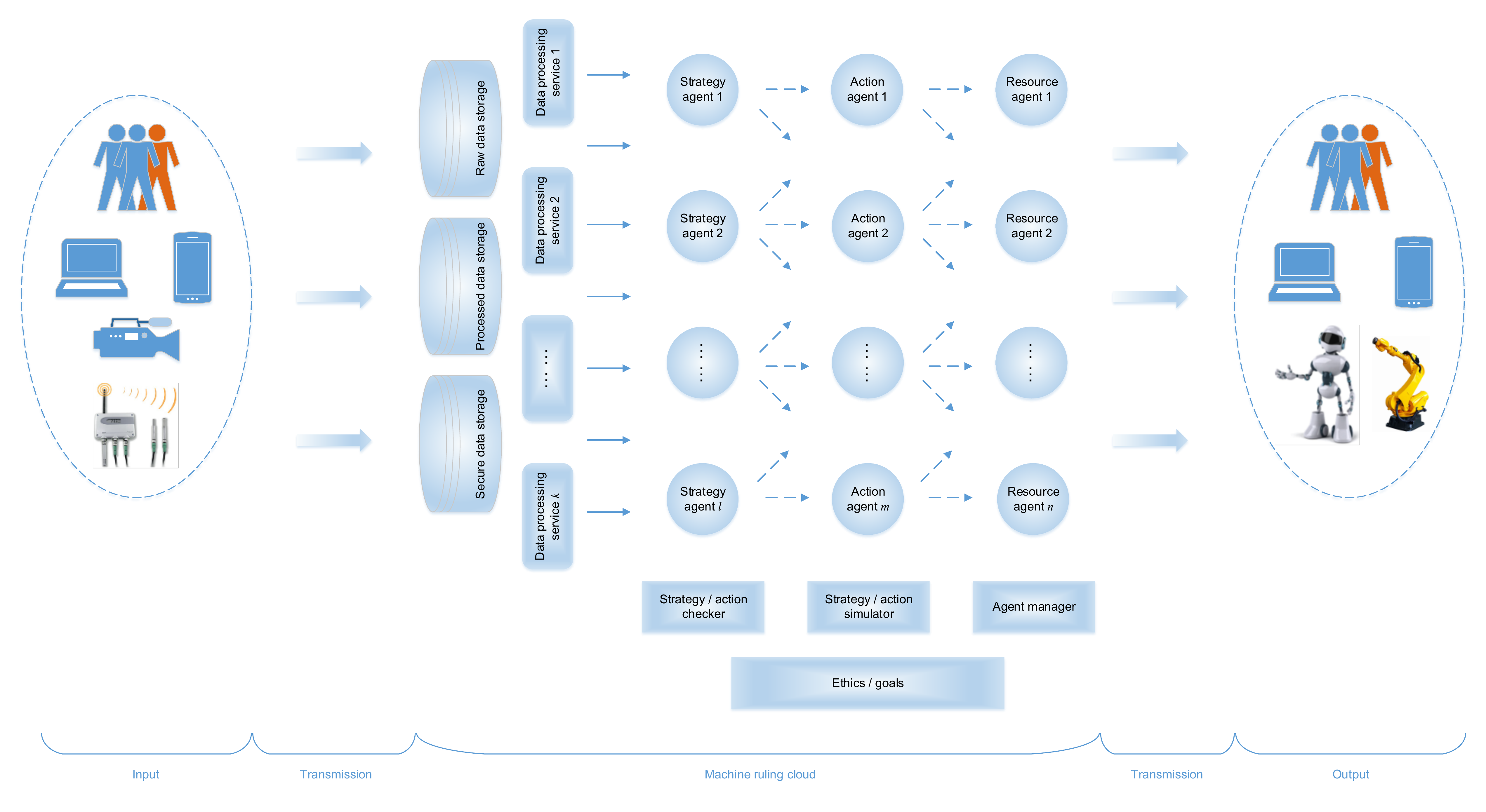}
	\caption{Proposed architecture of machine ruling.}
	\label{fig_architecture}
\end{figure*}

Based on the framework, an architecture of machine ruling is proposed, as shown in Fig.~\ref{fig_architecture}. Persons and objects in the dotted oval on the left generate inputs to the machine ruling cloud (which is the center part in Fig.~\ref{fig_architecture}). Inputs made by human beings, such as speech and gesture, 
or made by various objects and generalized sensors, such as temperature and velocity, 
are digitized and transmitted to the machine ruling cloud by either wireline (such as coaxial cable, power line, and optical fiber) or wireless (such as Wi-Fi, ZigBee, and 3G/4G). The received input data are stored in three data storage units, i.e., raw data storage, processed data storage, and secure data storage. The raw data storage stores unprocessed data, whereas the processed data storage keeps processed data or intermediate data. The secure data storage preserves privacy data and secure data, where data accesses are strictly limited.
The data processing service units perform elementary or common data processings, such as preprocessing, feature extraction, and decision making, providing subsequent agents with basic data processing supports. 

An agent is a virtual entity residing in the machine ruling cloud that comprises of a set of programs and devotes to a particular duty. Agents are classified into three types, i.e., strategy agents, action agents, and resource agents. Strategy agents, or policy agents, analyze the massive data in the data storage units with (or without) the supports provided by the data processing service units, and make implementable policies or specific plans. Each strategy agent specializes in one field or topic, e.g., environmental protection. It calls one or more action agents to implement the policies or carry out the plans. Action agents, or organization agents, organize projects, implement policies, or carry out plans. Each action agent fulfills a particular action, e.g., manufacturing electric vehicles, and interacts with one or more resource agents in organizing resources. Resource agents act as managers, brokers, or servants, in organizing resources, such as objects and persons. Every single object or person corresponds to one resource agent. A resource agent makes schedules and coordinates everything for the objects or persons within its limit. As an example, a resource agent arranges a person's daily schedule and assists the person in his or her daily life. Resource agents are also responsible for tracking and supervising persons for the purpose of accountability and compliance.

The outputs of the machine ruling cloud, including schedules for persons and commands for objects, 
are transmitted to the objects and facilities in the dotted oval on the right in Fig.~\ref{fig_architecture}. 
Once objects, e.g., robots, receive the output commands, they execute the commands right away. 
Note that the machine ruling cloud inputs and outputs in a real-time fashion.

There are a few more function modules on the bottom of Fig.~\ref{fig_architecture}. The strategy/action checker examines the outputs of each strategy agent and action agent to ensure full compliance with the ethics and goals provided by the ethics/goals module. It reserves the right to rectify or terminate any incompliant agent through the agent manager. The strategy/action simulator works with strategy agents and action agents and evaluates the effects and efficiencies of strategies, policies, plans, or actions by simulation. Once any of the strategies, policies, plans, or actions may result in unexpected situations, the strategy/action simulator informs the corresponding strategy agent or action agent to correct its outputs. The agent manager supervises and manages all the agents. It creates, adjusts, and destroys agents according to the needs of the strategy/action checker, the strategy/action simulator, and the ethics/goals module.
The ethics/goals module maintains and protects basic ethics and goals for the machine ruling cloud.
All the modules may be distributed in different data centers in the machine ruling cloud.

There are two phases for establishing the agents. In the first phase, all the strategy/action/resource agents are created and initialized by human beings. The ethics/goals module is also initialized by human beings. In the second phase, the machine ruling cloud takes over the management and automatically creates, adjusts, and destroys the agents, when relevant software and algorithms are ready.

\section{Enabling Technologies and Challenges}

As aforementioned, enabling technologies of machine ruling include but are not limited to Internet of things, mobile Internet, big data, cloud computing, and robotics.

In Internet of things, smart objects, usually embedded in physical objects and connected by networks to computers, make sense of their local situation and interact with human beings~\cite{iot1,iot2}. They sense, log, and interpret what is
occurring within themselves and the world, 
intercommunicate with each other, and exchange information with people~\cite{iot1}.
Although the definition of `things' has changed as technology evolved, the main idea of making computers sense information without the aid of human intervention remains the same~\cite{iot3}. In machine ruling, with Internet of things, various kinds of information about the physical world and human beings can be gathered by generalized sensors and transmitted to the machine ruling cloud.

Nowadays, people use mobile Internet in their daily lives.
With the rapid growth in the numbers of people owning smart mobile phones and the growth in 3G/4G/Wi-Fi  
subscriptions~\cite{mi}, 
more and more people are using smart mobile phones to access the Internet at an enhanced data rate.
Mobile Internet provides machine ruling with not only high-speed wireless accesses to the cloud, but also multipurpose input and output terminals, e.g., smart mobile phones, with multiple sensors.

Big data refers to those datasets whose size is beyond the ability of current 
software tools to capture, store, manage, and analyze. As technology advances over time, the size of datasets that qualify as big data also increases~\cite{bd}.
Architectures, methods, and algorithms for big data can also be applied to machine ruling, especially to the data processing service units and the agents shown in Fig.~\ref{fig_architecture}.

Cloud computing is a model for enabling ubiquitous and on-demand network access to a shared pool of configurable computing resources, such as computer servers, storage, and services, that can be rapidly provisioned and released with minimal service provider interaction~\cite{cc1}. 
Cloud computing refers to both the applications that are delivered as services over the Internet and the hardware and system software in the data centers providing those services~\cite{cc2}. 
The machine ruling cloud is built on cloud computing. 
With the support of hardware and system software of cloud computing, function modules of the machine ruling cloud can be implemented.

Speaking of robot, people may remind of electro-mechanical machines 
that are usually guided by computers. In machine ruling, robots are networked with the machine ruling cloud and employed as actuators or performers to execute instructions or fulfill tasks given by the machine ruling cloud, or to perform responding actions to interact with or to serve human beings.

Although these enabling technologies have laid the foundation of machine ruling, 
there is still a long way to go before machine ruling comes into reality. Technically speaking, 
the following challenges are worth being paid attention to and further investigated.

First, in sensing of information, how is all the various kinds of information from the physical world and human beings properly represented in digital forms? What information needs to be collected? And how it is collected? 

Second, in transmission of information, how to connect all the remote objects and human beings to the machine ruling cloud effectively and efficiently? How is the sensed digitized information transferred to the machine ruling cloud efficiently, reliably, and in real time? And how is the output information from the machine ruling cloud transferred to remote objects and human beings efficiently, reliably, and in real time?

Third, in data storage, what is the best way to organize and store all the massive data efficiently and safely? How to handle massive simultaneous data inquires?

Fourthly, in data processing, how to analyze and process the collected massive increasing time-variant data in time? 
How to handle the ultra-high-dimensional data collected from the physical world and human beings? How to find crucial clues and extract relevant core information from the massive data? 
How to apply the results of data processing to the agents? 

Fifthly, for the strategy/action/resource agents, how to decide the scope of 
each agent? How to realize and evolve ``correct'' ethics and goals? How to figure out and carry out plans? How to simulate and evaluate strategies, policies, and plans? 
How to ensure the instructions or commands outputted by the machine ruling cloud to be executed as expected? 

Sixthly, how to build and maintain a robust and secure machine ruling system?

With the advance of technology and further research, those challenges will be eventually solved.
The concept of machine ruling is expected to be gradually applied to daily life in the coming five to twenty years.

\section{Conclusion}
The concept, framework, and architecture of machine ruling have been proposed. The justifications, enabling technologies, and challenges for machine ruling have also been introduced. With the advance of relevant technologies and the increasing demands for inter-connected massive data processing as well as fair and efficient management and governance, machine ruling will eventually come into everyone's life. 

Methods and algorithms for the machine ruling cloud will be further investigated in the future.


%



\section*{Acknowledgment}

This work is supported by the Fundamental Research Funds for the Central Universities (N140404015).

\ifCLASSOPTIONcaptionsoff
  \newpage
\fi

\bibliographystyle{IEEEtran}

\bibliography{machine_ruling}

\end{document}